\documentclass[acmsmall,screen]{acmart}
\AtBeginDocument{%
  }

\usepackage{tabularx}  
\usepackage{booktabs}  
\usepackage{multirow}
\usepackage{booktabs,tabularx,multirow,caption}
\newcolumntype{Y}{>{\raggedright\arraybackslash}X} 
\usepackage[most]{tcolorbox} 
\tcbset{
  findingstyle/.style={
    enhanced, breakable,
    colback=black!5,      
    colframe=black,       
    boxrule=1.1pt,        
    arc=1mm,              
    left=2pt,right=2pt,top=1pt,bottom=1pt, 
  }
}

\newtcolorbox{findingbox}[1][]{findingstyle,#1}

\usepackage{tikz}
\usepackage{pgf-pie}
\usepackage{tikz}
\usetikzlibrary{arrows.meta,positioning,fit,shapes.misc}
\usepackage{enumitem}

\usepackage{pifont}
\usepackage{makecell} 
\usepackage[table]{xcolor} 
\newcommand{\cmark}{\textcolor{green!60!black}{\ding{51}}} 
\newcommand{\xmark}{\textcolor{red!80!black}{\ding{55}}}   

\begin{document}

\title{EvidenT: An Evidence-Preserving Framework for Iterative System-Level Package Repair}




\author{Chenyu Zhao}
\email{zhaochenyu@mail.nankai.edu.cn}
\affiliation{%
  \institution{Nankai University}
  \city{Tianjin}
  \country{China}
}

\author{Minghua Ma}
\authornote{Corresponding author.}
\email{minghuama@microsoft.com}
\affiliation{%
  \institution{Microsoft}
  \city{Seattle}
  \state{WA}
  \country{USA}
}

\author{Shenglin Zhang}
\email{zhangsl@nankai.edu.cn}
\affiliation{%
  \department{College of Software}
  \institution{Nankai University}
  \city{Tianjin}
  \country{China}
}
\additionalaffiliation{%
  \institution{Haihe Laboratory of Information Technology Application Innovation}
  \city{Tianjin}
  \country{China}
}
\additionalaffiliation{%
  \institution{Key Laboratory of Data and Intelligent System Security, Ministry of Education}
  \city{Tianjin}
  \country{China}
}

\author{Zeshun Huang}
\email{2213900@mail.nankai.edu.cn}
\affiliation{%
  \institution{Nankai University}
  \city{Tianjin}
  \country{China}
}

\author{Yongqian Sun}
\email{sunyongqian@nankai.edu.cn}
\affiliation{%
  \department{College of Software}
  \institution{Nankai University}
  \city{Tianjin}
  \country{China}
}
\additionalaffiliation{%
  \institution{Tianjin Key Laboratory of Software Experience and Human Computer Interaction}
  \city{Tianjin}
  \country{China}
}

\author{Chetan Bansal}
\email{chetanb@microsoft.com}
\affiliation{%
  \institution{Microsoft}
  \city{Redmond}
  \state{WA}
  \country{USA}
}

\author{Saravan Rajmohan}
\email{saravan.rajmohan@microsoft.com}
\affiliation{%
  \institution{Microsoft}
  \city{Redmond}
  \state{WA}
  \country{USA}
}

\author{Dan Pei}
\email{peidan@tsinghua.edu.cn}
\affiliation{%
  \institution{Tsinghua University}
  \city{Beijing}
  \country{China}
}

\renewcommand{\shortauthors}{Zhao et al.}






\renewcommand{\shortauthors}{Trovato et al.}
\def\eg{e.g.,~}
\def\etal{et al.~}
\def\name{\textsc{EvidenT}}

\begin{abstract}
Frequent toolchain updates and the expanding diversity of instruction set architectures (ISAs) have made large-scale system-level software package repair a critical task.
Diagnosing and repairing build failures remains challenging due to heterogeneous failure evidence, complex dependency constraints, and architecture-specific build conventions.
While recent LLM-based repair methods have shown promise for project-level source code fixes, they struggle with system-level repair where failures involve multi-language artifacts (\eg build recipes, scripts, and source archives) and require iterative validation through external build services.
In this paper, we first conduct a systematic empirical study of real-world system-level build failures.
Our findings reveal that 72\% of failures stem from dependency and environment misconfigurations rather than isolated code defects, suggesting that effective repair must prioritize packaging logic and iterative feedback.
Motivated by these insights, we propose \name{}, an evidence-preserving repair framework that decouples iteration-aware evidence management from tool execution. 
\name{} comprises (1) an external Build Service for reproducible build execution and feedback; (2) an Evidence-Preserving Repair Controller that performs cross-modal fusion of repair history, knowledge context, and build artifacts; and (3) an automated Repair Orchestrator that executes a suite of modular tools for failure localization and system-level repair actions within a closed-loop validation environment.
We evaluate \name{} on a benchmark of 219 real-world RISC-V package build failures.
\name{} successfully repairs 118 packages (53.88\%), substantially outperforming state-of-the-art agentic baselines (20.55\%) and direct LLM-based repair (1.83\%).
To demonstrate its architectural generality, we extend \name{} to legacy ISAs by updating only ISA-specific knowledge context. 
In preliminary experiments, it achieves success rates of 41.77\% on aarch64 and 46.99\% on x86\_64, showcasing its robustness across diverse hardware ecosystems.

\end{abstract}




\maketitle

\section{Introduction}

As software ecosystems continue to evolve, large-scale system-level package builds triggered by toolchain updates, system component upgrades, and architecture evolution have become routine.
Despite extensive automation, build failures remain inevitable in practice due to complex dependency graphs, architecture-specific constraints, and fragmented \emph{failure evidence} scattered across build stages.
Once such failures occur, repair is often labor-intensive and costly, delaying software releases and incurring substantial maintenance effort.
For example, Apple’s transition from Intel x86 to ARM-based M1 processors required rebuilding thousands of packages, many of which encountered compatibility problems that took months to resolve~\cite{ford2021migrating}. 

As summarized in Table~\ref{tab:capability-matrix}, existing approaches exhibit significant limitations when applied to system-level packages. 
Traditional approaches~\cite{DBLP:conf/icse/HassanW18, DBLP:journals/infsof/MakC24, DBLP:conf/issta/FanWW0SZ20, DBLP:journals/tosem/NourryKSSK25, DBLP:conf/kbse/Zhang0H0Z22, DBLP:conf/icse/HenkelSTdR21, DBLP:conf/icse/Sun25} primarily rely on handcrafted rules, historical repair patterns, or specific heuristics tailored to isolated failure modes.
They typically fail to generalize to real-world package ecosystems characterized by multi-language environments, heterogeneous-artifacts (\eg packaging recipes, build scripts, and source archives), and the need for end-to-end build validation.

While Large Language Models (LLMs) have been explored as automated assistants for log analysis and software debugging~\cite{10.1145/3460345,10771398}, our evaluation on 219 failed packages from the Open Build Service (OBS)\footnote{OBS is an open build system that supports package building: \url{https://openbuildservice.org/}} reveals a clear limitation: 
LLMs successfully repaired only 4 cases when directly analyzing build logs without external tool support. 
This suggests that LLMs, in isolation, cannot handle the complexity of system-level package repair. 
Recent agent-based extensions~\cite{RepairAgent_2025, CXXCrafter_2025, agentless_2025_fse} improve automation through multi-step tool use.
Nevertheless, they typically designed for source-centric or single-artifact environments and still lack explicit mechanisms to preserve and reuse failure evidence across iterations.
Consequently, these agents frequently repeat ineffective actions and fail to converge under heterogeneous and architecture-specific conditions.

\begin{table*}[t]
  \centering
  \captionsetup{font=small}
  \scriptsize
  \caption{Capability matrix of related approaches compared to \name{}. "Category" indicates the methodology class, and checkboxes denote supported features essential for system-level package repair.}
  \label{tab:capability-matrix}
  \begin{tabularx}{\textwidth}{l X c c c c}
    \toprule
    \textbf{Category} & \textbf{Method} & \makecell{\textbf{Multi-}\\\textbf{language}} & \makecell{\textbf{Heterogeneous-artifact}\\\textbf{Pipeline}} & \makecell{\textbf{Automatic}\\\textbf{Repair}} & \makecell{\textbf{System-level}\\\textbf{Build}} \\
    \midrule
    
    \multirow{7}{*}{\makecell[l]{Traditional\\Approaches}} 
      & HireBuild / History-Driven Fixing~\cite{DBLP:conf/icse/HassanW18} & \xmark & \xmark & \cmark & \xmark \\
      & Escaping Dependency Hell~\cite{DBLP:conf/issta/FanWW0SZ20}          & \cmark & \xmark & \xmark & \xmark \\
      & My Fuzzers Won’t Build~\cite{DBLP:journals/tosem/NourryKSSK25}            & \cmark & \xmark & \xmark & \xmark \\
      & BuildSonic~\cite{DBLP:conf/kbse/Zhang0H0Z22}                        & \cmark & \xmark & \xmark & \xmark \\
      & Automatic Build Repair~\cite{DBLP:journals/infsof/MakC24}            & \xmark & \xmark & \cmark & \xmark \\
      & Shipwright~\cite{DBLP:conf/icse/HenkelSTdR21}                                & \cmark & \xmark & \xmark & \xmark \\
      & Intelligent Automation~\cite{DBLP:conf/icse/Sun25}                      & \xmark & \xmark & \cmark & \xmark \\
    \midrule
    
    \multirow{3}{*}{\makecell[l]{Agent-based\\Approaches}} 
      & RepairAgent~\cite{RepairAgent_2025}                        & \xmark & \xmark & \cmark & \xmark \\
      & CXXCrafter~\cite{CXXCrafter_2025}                        & \xmark & \xmark & \cmark & \xmark \\
      & Agentless~\cite{agentless_2025_fse}                         & \xmark & \xmark & \cmark & \xmark \\
    \midrule
    
    \textbf{Ours} & \textbf{\name{}} & \textbf{\cmark} & \textbf{\cmark} & \textbf{\cmark} & \textbf{\cmark} \\
    \bottomrule
  \end{tabularx}
\end{table*}

Our empirical study in Section~\ref{sections/empirical_study} demonstrates that system-level package failures are highly diverse. 
Specifically, 56\% of failures stem from dependency, compilation, or packaging issues, while the remaining 44\% occur during testing and validation.
Furthermore, among successful repairs, 72\% primarily involve adjustments to build configurations, dependency, or environment settings, and 28\% require modifications to source archives.
These findings indicate that effective automated repair cannot rely solely on isolated log inspections or source-centric reasoning.


In practice, system-level package repair produces rich \emph{failure evidence}, including build feedback, dependency constraints, package structure, and prior repair outcomes.
However, such evidence is often fragmented, noisy, and iteration-dependent, making it difficult to exploit effectively in automated repair.
Based on our empirical observations, we distill three fundamental challenges:

\noindent\textbf{Challenge 1: How to acquire actionable failure evidence.} 
Failure evidence is submerged within voluminous build logs, complex dependency resolution outputs, and heterogeneous package artifacts.
Extracting precise repair guidance from such noisy data, while compensating for the agent's lack of architecture-specific experience through a knowledge context, remains a primary obstacle.

\noindent\textbf{Challenge 2: How to organize and fuse heterogeneous evidence.}
Evidence signals span multiple modalities, such as semi-structured text (logs), graph structures (dependency constraints), and cached findings from multiple tools. 
Organizing these heterogeneous evidence streams into a unified and structured representation is essential for guiding artifact-level repair decisions.

\noindent\textbf{Challenge 3: How to preserve and reuse iteration-dependent evidence.}
System-level package repair is an iterative process of analysis, modification, and validation. 
Most existing methods treat each attempt in isolation, failing to preserve evidence across iterations. 
Without explicit evidence preservation, agents often repeat ineffective actions
and explore redundant repair paths, leading to unstable repair loops.

These challenges motivate the design of a repair framework that can coordinate multiple tools and explicitly preserve failure evidence across repair iterations.
To this end, we propose \textbf{\name{}}, an evidence-preserving framework for iterative system-level package repair.
\name{} integrates analysis, repair, and validation into a unified workflow, in which repair outcomes are continuously verified through external build services such as OBS and Docker.
To support scalable and decoupled tool orchestration, \name{} adopts the Model Context Protocol (MCP)~\cite{hou2025modelcontextprotocolmcp} as a standardized substrate for tool interaction.

Architecturally, \name{} comprises three main components.
To distill actionable failure evidence from noisy build signals and to complement missing expertise with auxiliary knowledge (Challenge~1), we design an \emph{Analysis and Repair Orchestration} module that coordinates specialized tools for failure localization, artifact inspection, and system-level repair actions across packaging recipes, archives, and source code.
To unify and structurally fuse heterogeneous evidence streams (Challenge~2), we introduce an \emph{Evidence-Preserving Repair Controller} that maintains an iteration-aware evidence context and injects fused evidence into fixed prompt slots for disciplined repair decisions.
To preserve an iteration-aware evidence (Challenge~3), we incorporate a \emph{Build-based Validation and Feedback Loop} that executes patched packages in reproducible external build environments and feeds build outcomes back to the controller as new failure evidence, preventing redundant exploration and supporting stable convergence.

In summary, this paper establishes evidence preservation as a necessary principle for scalable iterative repair of system-level packages, and makes the following contributions:
\begin{itemize}

\item \textbf{The first empirical characterization of system-level repair.}
We present the first large-scale empirical study of real-world system-level package build failures, analyzing 219 failed packages from OBS and further examining 100 representative cases in depth to characterize root causes, repair targets, and failure stages across heterogeneous artifacts.

\item \textbf{Evidence-preserving repair framework.}
We propose \name{}, an evidence-preserving framework that coordinates heterogeneous analysis,
repair, and build-validation tools in a unified iterative workflow, enabling automated repair
beyond source-centric and single-artifact settings.

\item \textbf{Superior performance and generalization.}
\name{} achieves repair success rates of 53.88\% on RISC-V~\cite{cui2023risc}, 41.77\% on aarch64, and 46.99\% on x86\_64 OBS package failures, outperforming LLM-only repair and representative agent-based baselines.

\item \textbf{Open-source implementation and datasets.}
We release our implementation and datasets through an anonymous repository to support
reproducibility and future research\footnote{\url{https://anonymous.4open.science/r/EvidenT-1638/README.md}}.

\end{itemize}

\section{Background \& Related Work}

\subsection{Package Building}

\textbf{Overview.}
Transforming source code and build specifications (\eg spec files) into installable binary packages is a core task in software maintenance and distribution. 
Traditional build tools (\eg \texttt{rpmbuild}, \texttt{dpkg-buildpackage}) as well as integrated platforms such as the Open Build Service (OBS) and Koji \cite{koji_definition} provide automation in dependency resolution, compilation, and packaging\cite{Aidasso2025, Macho2024, Gibb2025}. Container-based solutions (\eg Docker) further improve reproducibility by encapsulating the build environment \cite{Cito2016, Matelsky2018, Moreau2021, Vessel2025}. 
While containerized approaches mitigate environment inconsistencies, they do not eliminate other sources of build failures arising from package configuration, dependencies, or code-level issues. 
Consequently, builds still fail frequently due to a wide range of factors spanning
dependencies, source code, and build specifications~\cite{DBLP:conf/icse/HassanW18, icse2021codereview, icse2019social}.

\noindent\textbf{Challenges in Large-Scale, Multi-Language Builds.}
Large-scale system-level package building fundamentally differs from project-level builds in both scope and complexity.
Packages often combine multiple programming languages, rely on heterogeneous build systems, and must be validated across diverse architectures.
Different ecosystems (\eg C/C++, Java, Python) rely on distinct build systems and dependency management tools~\cite{pip2025,maven2025,cmake2025}, which complicates automated repair beyond single-language scenarios studied in prior work~\cite{hassan2017autobuild, wang2023pyconflict, CXXCrafter_2025}. 
Existing studies typically address only one dimension of this complexity.
Java-focused work shows that some failures can be mitigated by modifying build scripts or configuration files~\cite{hassan2017autobuild}.
Python-oriented research emphasizes dependency conflict resolution within package managers such as pip and conda~\cite{wang2023pyconflict}.
For C/C++ projects, CXXCrafter~\cite{CXXCrafter_2025} reports that only a small fraction of projects can be automatically built without substantial human intervention.
Other research has explored fast build triage and the impact of build system evolution \cite{ase2021buildfast, ase2022changes}, but these techniques focus on diagnosis rather than automated repair.
Overall, existing studies address individual aspects of the problem, while comprehensive automated repair for large-scale, multi-language, system-level package builds remains an open challenge.

\subsection{Agent and MCP}

\noindent\textbf{LLM-Based Agents.}
Large language models (LLMs) have achieved notable success in knowledge storage~\cite{zhang2024distilling}, code generation~\cite{huning2021automatic}, and semantic understanding~\cite{tang2023semantic}. 
LLM-based agents integrate these models with external tools, planning mechanisms, and environment interaction, enabling more autonomous execution~\cite{liu2023agentbench, ibm2023agents, barua2024autonomous, shetty2024buildingagents}. 
In software engineering, agent-based systems have been applied to automated debugging, program repair, and build-related tasks~\cite{chen2024rcacopilot, issre2024triage, fse2024monitor}.
Representative examples include RepairAgent~\cite{RepairAgent_2025}, CXXCrafter~\cite{CXXCrafter_2025}, and Agentless~\cite{agentless_2025_fse}, which demonstrate that LLM-driven agents can automate project-level repairs or build fixes within constrained benchmarks or single-language settings.
However, these agents often assume that failures can be localized within a limited set of source artifacts or a single language ecosystem.
They lack native support for coordinated manipulation of heterogeneous artifacts such as spec files, source archives, and build metadata, and do not explicitly preserve or reason over failure evidence across repair iterations.
Consequently, their applicability to system-level package repair remains limited.

\noindent\textbf{Model Context Protocol (MCP).}
The Model Context Protocol (MCP) standardizes interactions between LLMs and external tools by providing a unified interface for tool invocation and stateful communication.
MCP supports cross-model interoperability~\cite{krishnan2025mcp}, modular and reusable tool integration, dynamic discovery of tools at runtime~\cite{shahid2025llmtoolcalling}, and structured state exchange via JSON-RPC.
Importantly, MCP enables the preservation and reuse of contextual information such as long build logs and intermediate tool outputs across interactions~\cite{krishnan2025mcp}.
These properties make MCP a suitable substrate for implementing iterative, evidence-aware workflows in complex tool-based settings.
Nevertheless, MCP itself is domain-agnostic and does not address build-specific challenges such as failure localization across artifacts or architecture-aware reasoning.
A dedicated repair framework is still required to leverage MCP effectively in system-level package repair.

\section{Empirical Study}
\label{sections/empirical_study}
%
%
%

In this section, we conduct an empirical study to characterize real-world system-level package build failures and to understand the fundamental challenges that hinder automated repair.
Using build failures on the RISC-V ISA as a representative setting, we analyze failure causes, repair difficulty, and cross-artifact modification patterns to derive actionable guidance for the design of an automated system-level repair framework.
The research questions, dataset, and key findings are structured as follows.

\noindent\textbf{Research Questions.} 
We formulate the following research questions to understand the root causes of system-level
build failures and the practical challenges that limit automated repair:
\begin{itemize}[leftmargin=*]
    \item RQ1 (Failure Causes).
    What are the primary root causes of build failures observed among these packages?
    
    \item RQ2 (Repair Difficulty).
    Why is repairing failed packages challenging, and how do failed packages differ from their nearest successful builds in terms of repair locations and required modifications across artifacts? 
\end{itemize}

\noindent\textbf{Dataset.} 
We collect the 100 most recent failed system-level packages on RISC-V from the Open Build Service (OBS), together with their complete build logs and accessible source directories.
These packages represent a subset of the final evaluation dataset and are selected to
separate empirical characterization from framework evaluation.

To support RQ2, for each failed build we additionally retrieve the temporally nearest \emph{successful} build of the same package (by build date) and treat it as a control group for comparative analysis. 
To ensure validity and reproducibility, we apply three inclusion criteria:
(1) each failed build is followed by at least one subsequent successful build, enabling controlled comparison between failure and success states;
(2) complete package artifacts and corresponding failure logs are accessible for diagnosis (\eg spec files, sources, and auxiliary packaging files);
and (3) both the failure and success builds can be reliably reproduced using the original source code, which eliminates inconsistencies from environmental variables.

The selected packages cover diverse technical domains, encompassing development toolchains, cloud-native and container management, network services, security and cryptography, graphics and multimedia, system monitoring, data analysis and programming support, hardware adaptation, Python ecosystem and libraries, and command-line tools.
We analyze the source code language distribution of the 100 system software packages. 
Python dominates the dataset (48\%), followed by C/C++ (17\%) and Go (10\%). 
Smaller portions are implemented in Rust (3\%), Ruby (3\%), and Perl (2\%), while 17\% adopt mixed-language implementations. 
For instance, \texttt{apache2-mod\_wsgi}~\footnote{\url{https://build.opensuse.org/package/show/openSUSE:Factory/apache2-mod_wsgi}} combines extension modules in C with application logic in Python, reflecting the cross-language characteristics of certain system-level software packages and explains why agents designed for specific language environments cannot be directly transferred, posing a challenge to our work.

\subsection{Failure Causes from Build Logs (RQ1)}
To answer RQ1, we manually analyze and classify the build failure logs of the 100 software packages.
Two software engineers independently conduct fine-grained log inspection and root-cause labeling using domain knowledge of Linux packaging and build systems; disagreements are resolved through discussion to produce a unified taxonomy.
Ultimately, as shown in Table~\ref{tab:build-failure-classification}, four primary causes of build failures are identified: \emph{Dependency Failures}, \emph{Compilation Failures}, \emph{Test Failures}, and \emph{Packaging Failures}.
Detailed descriptions of these categories are provided below.

\begin{table*}[t] 
  \centering
  \captionsetup{font=small}
  \scriptsize
  \caption{Classification of Build Failure Cases. "Category" refers to the expert-defined failure category, "Subcategory" is the further division based on the failure, and "Count" indicates the number of packages in each sub-category.}
  \label{tab:build-failure-classification}
  \begin{tabularx}{\textwidth}{l l X c}
    \toprule
    \textbf{Category} & \textbf{Subcategory} & \textbf{Description} & \textbf{Count} \\
    \midrule
    \multirow{2}{*}{Dependency Failures}
      & Missing Dependency   & Missing versions of required components, libraries, and modules in the build environment. & 15 \\
      & Dependency Conflict  & Version mismatch or reliance on deprecated dependencies. &  6 \\
      \cmidrule{2-4}
      & \textbf{Sum}    &     & \textbf{21} \\
    \midrule
    \multirow{3}{*}{Compilation Failures}
      & Code-Level Issues      & Errors related to syntax, type definitions, and logical inconsistencies in the source code. & 11 \\
      & Build Errors           & Issues with build scripts, compiler configurations, and compilation processes. & 11 \\
      & API \& Function Issues & Calling deprecated, removed, or incompatible functions and APIs. &  5 \\
      \cmidrule{2-4}
      & \textbf{Sum}      &      & \textbf{27} \\
    \midrule
    \multirow{3}{*}{Test Failures}
      & Code \& Compatibility Issues    & Outdated, incompatible, or incorrect test code.   & 11 \\
      & Assertion \& Compliance Violations & Violation of test assertions or compliance checks.  & 17 \\
      & Test Execution Problems  & Failures related to the execution of the test framework itself. & 16 \\
      \cmidrule{2-4}
      & \textbf{Sum}     &      & \textbf{44} \\ 
    \midrule
    \multirow{2}{*}{Packaging Failures}
      & Packaging Configuration Errors   & Incorrect settings or syntax in the packaging specification and configuration files.  &  6 \\
      & Installation/Verification Failures & Problems occurring during the actual installation or final verification stages. &  2 \\
      \cmidrule{2-4} 
      & \textbf{Sum}         &           & \textbf{8}\\
    \midrule
    \textbf{Total} &  & &  \textbf{100} \\
    \bottomrule
  \end{tabularx}
\end{table*}

\noindent\textbf{Dependency Failures (21/100).}
Dependency failures arise when required external components (\eg libraries, tools, or modules) are missing, misconfigured, or incompatible with the build environment.
We observe 21/100 cases in this category, including \emph{Missing Dependencies} (15) and \emph{Dependency Conflicts} (6).
Representative examples include missing \texttt{wlroots} in Meson projects, absent \texttt{cmake} required by the Rust crate \texttt{aws-lc-sys}, missing Python modules such as \texttt{pkg\_resources}, and version incompatibilities such as PyO3 under Python~3.13 or \texttt{quic-go} under Go~1.21.
Comparing with successful builds suggests that fixes often require adjusting \texttt{BuildRequires}/runtime dependencies, exposing dependencies via pkg-config/CMake, upgrading incompatible versions, or providing missing tools/headers.

\noindent\textbf{Compilation Failures (27/100).}
Compilation failures occur when compilers reject source code or build instructions.
We observe 27/100 cases, including \emph{Code-level Issues} (11; \eg missing headers, incomplete types, invalid callback signatures), \emph{Build Errors} (11; \eg invalid flags, unsupported options, or toolchain constraints such as LTO back-end failures under specific RISC-V vector configurations), and \emph{API \& Function Issues} (5; \eg deprecated OpenSSL~3.0 APIs promoted to errors under \texttt{-Werror}, or removed C APIs in Python~3.13).
Typical fixes include modernizing API usage, adding headers/feature guards, correcting compiler flags, or aligning code with updated dependency versions.

\noindent\textbf{Test Failures (44/100).}
Test failures occur during validation (\texttt{\%check}) when test suites fail.
We observe 44/100 cases, including \emph{Code \& Compatibility Issues} (\eg deprecated unittest methods, removed modules such as \texttt{imp} in Python~3.12, type mismatches), \emph{Assertion \& Compliance Violations} (\eg incorrect assertions such as \texttt{assertEquals}, mismatched AST/XML outputs, compliance warnings treated as errors), and \emph{Test Execution Problems} (\eg misconfigured discovery, missing CTest binaries, GUI-induced segmentation faults).
Fixes typically require updating assertions/APIs, correcting discovery paths and environment variables, and provisioning test-only dependencies to stabilize harness execution.

\noindent\textbf{Packaging Failures (8/100).}
Packaging failures originate from packaging metadata or scripts.
We observe 8/100 cases, including \emph{Packaging Configuration Errors} (6; \eg macro/syntax issues, phase ordering, install path misconfigurations such as \texttt{libtool} refusing installs outside \texttt{/usr}) and \emph{Installation/Verification Failures} (2; \eg file list mismatches such as ``glob not found'', or cross-subpackage file conflicts).
These failures are commonly addressed by fixing spec macros/phase ordering, repairing install paths, and adding missing \texttt{BuildRequires}.

\begin{findingbox}
\textbf{Finding 1:} Build failures span multiple stages: \textbf{56\%} arise from dependency/compilation/packaging issues, while \textbf{44\%} are test-related. 
Each category corresponds to distinct failure evidence and repair locations, making accurate root-cause analysis essential for automated repair.
\end{findingbox}

\noindent\textbf{Implication.}
Failure categories correspond to orthogonal repair actions and target artifacts.
Without explicit \emph{Root Cause Analysis (RCA)}, automated repair easily degenerates into trial-and-error over irrelevant files and log fragments.
This motivates an RCA tool suite with two core functions:
(1) extracting and prioritizing anomalous log segments that are most likely causal, and
(2) localizing candidate artifacts for modification (\eg spec files, manifests, source/header files, and test scripts).

\begin{table}
  \centering
  \captionsetup{font=small}
  \scriptsize
  \caption{Types of Repair Observed in 100 Failed Packages. "Category" refers to expert-defined modification classes and "Modification details" show the concrete operations required for package repair}
  \label{tab:modification-types}
  \begin{tabular}{l p{0.46\textwidth} l r}
    \toprule
    \textbf{Category} & \textbf{Description} & \textbf{Modification details} & \textbf{Count} \\
    \midrule
    \multirow{4}{*}{Spec Modification} 
      & \multirow{4}{=}{Complex configuration adjustments, such as correcting build macro definitions or updating dependency declarations.}
      &  modified lines $\leq 50$ & 45 \\
      &  & modified lines in $51$--$100$ & 23 \\
      &  & modified lines $> 100$ & 4 \\
      \cmidrule{3-4}
      & & \textbf{Sum} & \textbf{72} \\
    \midrule
    \multirow{4}{*}{Packaging-Related Changes}
      & \multirow{4}{=}{Vendor archives, service/metadata files, auxiliary data.}
      & Added files & 9 \\
      & & Deleted files & 2 \\
      & & Modified files & 17 \\
      \cmidrule{3-4}
      & & \textbf{Sum} & \textbf{28} \\
      \bottomrule
  \end{tabular}
\end{table}

\subsection{Repair Locations and Modification Complexity (RQ2)}
To answer RQ2, we compare each failed package against its nearest successful build.
We design an automated workflow that extracts, aligns, and cross-validates differences across four dimensions:
(1) directory structure, (2) source files, (3) specification (spec) files, and (4) packaging-related archives and auxiliary files.
Specifically, the analysis is performed at both the file and line levels, allowing us to capture fine-grained code edits, file insertions/deletions, and complex specification adjustments.
Applying this workflow to 100 paired builds yields the modification distribution in Table~\ref{tab:modification-types}, with two dominant categories.

\noindent\textbf{Spec Modifications (72/100).}
Spec files govern build phases and dependency declarations.
Most successful repairs require configuration-level adjustments such as macro/flag corrections, dependency updates, and patch instruction changes.
Quantitatively, most spec modifications are relatively small in scale, with $\leq50$ modified lines (45 cases). 
However, a non-trivial fraction required medium-scale edits of 51–100 lines (23 cases), and a smaller but significant set involved large-scale rewrites exceeding 100 lines (4 cases). 
This prevalence indicates that automated repair must prioritize reasoning over build
specifications and dependency configurations, rather than focusing solely on source code.

\noindent\textbf{Packaging-Related Changes (28/100).}
Beyond spec file modifications, a notable subset of repairs targeted the source archives or auxiliary files. 
These interventions generally fall into three categories:
(1) \emph{File additions} (9 cases): new files are introduced to address missing components, provide necessary datasets, or integrate third-party libraries absent from the original package; 
(2) \emph{File deletions} (2 cases): obsolete or conflicting files are removed to prevent incompatibilities, such as legacy helper scripts or outdated binaries that disrupted the build process; and 
(3) \emph{File modifications} (17 cases): existing scripts, metadata, or auxiliary configurations are revised to satisfy system-level requirements. 
Typical adjustments included updating \texttt{.service} files, refining \texttt{systemd} settings, and correcting metadata inconsistencies.


\begin{findingbox}
\textbf{Finding 2:} 
In \textbf{72\%} of cases, successful repair primarily involves spec-level configuration and
dependency/environment adjustments, while the remaining \textbf{28\%} require changes to
archives or auxiliary package artifacts beyond the spec file.
\end{findingbox}

\noindent\textbf{Implication.}
This distribution shows that effective automated repair must reason across heterogeneous
artifacts, rather than treating build failures as isolated source-code defects.
Spec-level modifications often require non-trivial reasoning over build environments,
dependency specifications, and architecture-specific configurations, even when changes are
relatively localized.
In contrast, packaging-related repairs demand accurate comprehension of source archives and
packaging conventions, where insufficient structural understanding can lead to incorrect or
incomplete modifications.

\section{\name{}}
%
%
%



In view of the challenges identified in Section~\ref{sections/empirical_study}, we design
\name{}, an evidence-preserving framework for automated system-level package repair.
\name{} targets build failures arising from heterogeneous artifacts, distributed failure
signals, and iterative repair dependencies that commonly occur in large-scale package ecosystems.

\subsection{Overview}

\begin{figure}
  \centering
  \includegraphics[width=\linewidth]{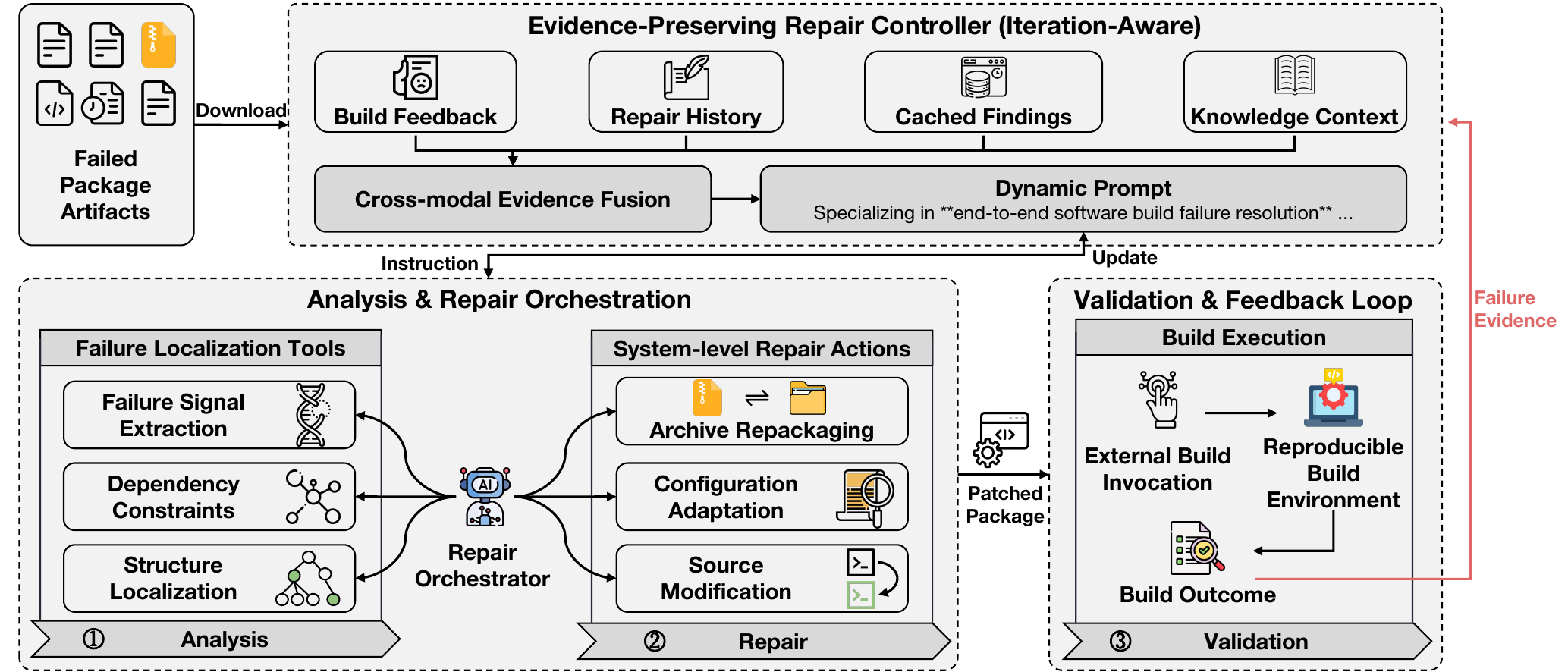}
  \caption{The framework of \name{}. The evidence-preserving repair controller maintains iteration-aware failure evidence, while the tool orchestrator exposes analysis, repair, and validation tools.
  }
\label{fig:framework}
\end{figure}

Figure~\ref{fig:framework} illustrates the overall architecture of \name{}.
The framework separates evidence tracking from tool execution and comprises three components:
(1) an external \emph{Build Service} that provides reproducible build results and concrete failure logs,
(2) an \emph{Evidence-Preserving Repair Controller} that maintains iteration-aware repair context
and constructs a dynamic prompt to guide repair actions, and
(3) an MCP-based tool layer that executes \emph{Analysis and Repair Orchestration} and
\emph{Validation and Feedback Loop}.
Together, these components form an iterative loop that analyzes failures, applies targeted
repairs, and validates outcomes through real builds.

\subsection{Evidence-Preserving Repair Controller}
\label{sec:context-manager}
The Evidence-Preserving Repair Controller maintains an iteration-aware evidence context,
conditioning each repair iteration on the latest build feedback, prior edits, cached analysis results, and relevant domain knowledge.
This design prevents repeated ineffective modifications and enables systematic, evidence-driven repair.

\subsubsection{Evidence Components}
\label{sec:evidence_components}
\name{} organizes the evidence context into four complementary components.
Together, these four components answer:
(1) what failed (Build Feedback),
(2) what has been tried (Repair History),
(3) what we know about the current package state (Cached Findings),
and (4) what similar failures looked like before (Knowledge Context).

\noindent\textbf{Build Feedback.}
To ensure iteration-aware failure analysis grounded in observable build behavior, \name{} captures iteration-level failure evidence produced by the external build service.
In \name{}, each repair iteration is validated through a clean rebuild on the Open Build Service (OBS), yielding
(i) the complete build log and (ii) a structured build outcome (\eg succeeded, failed, or unresolvable).
Only the most recent build feedback is retained.
This design enforces a strict iteration boundary, ensuring that failure analysis is always grounded in the latest observed behavior and prevents outdated logs from misleading later iterations.
The build feedback serves as the primary diagnostic signal for subsequent failure localization and repair planning.

\noindent\textbf{Repair History.}
To avoid repeated ineffective or contradictory repair actions across iterations, \name{} explicitly maintains a repair history that records all artifact-level modifications applied in previous iterations.
Each history entry captures the modified file path together with its updated content.
Unsuccessful repair actions are preserved as negative evidence and explicitly exposed to the LLM in subsequent iterations.
By conditioning repair decisions on accumulated modification history, \name{} enables progressive refinement of repair strategies.

\noindent\textbf{Cached Findings.}
To ensure the consistency of multi-step reasoning, \name{} caches the outputs of deterministic analysis tools, such as directory structure summaries and parsed packaging recipes. 
These results remain invariant unless the underlying artifacts are modified within the same iteration. 
This mechanism minimizes redundant tool invocations and token overhead, thereby maintaining a stable context for reasoning. 
Cached entries are refreshed at the beginning of each iteration to reflect any artifact updates
introduced by the previous repair attempt.

\noindent\textbf{Knowledge Context.}
To provide auxiliary evidence for architecture-aware and ecosystem-specific reasoning, \name{} maintains a knowledge context.
It aggregates curated knowledge derived from historical issue discussions, pull-request reviews, and architecture-specific documentation related to system-level package failures.
This knowledge is retrieved on demand and complements build-derived evidence by encoding recurring failure patterns, toolchain conventions, and ISA-specific pitfalls that are not explicitly exposed in build logs.
By integrating such contextual evidence, \name{} enables informed repair decisions beyond isolated source-level analysis.

\subsubsection{Cross-modal Evidence Fusion}
\label{sec:evidence_fusion}
To transform heterogeneous repair signals into actionable instructions, \name{} performs \emph{cross-modal evidence fusion}. 
Rather than naively concatenating data, the system integrates build feedback, repair history, cached findings, and knowledge context into a unified evidence representation.

The fusion process enforces three principles.
First, \emph{temporal grounding} ensures strict iteration boundaries by retaining only the failure evidence produced by the most recent rebuild.
Second, a \emph{slot-based prompt structure} organizes the four evidence components into a fixed, non-overlapping schema. 
Instead of interleaving raw snippets, \name{} (i) distills \emph{Build Feedback} into a concise set of distilled failure signals and the current build outcome; 
(ii) represents \emph{Repair History} as an ordered list of prior edits alongside their validation results; 
(iii) normalizes \emph{Cached Findings} into structured key summaries with stable identifiers; and (iv) injects \emph{Knowledge Context} only when it is relevant to the current failure. 
This structured organization makes the prompt explicitly iteration-aware while reducing ambiguity caused by redundant or inconsistent evidence sources.
Third, \emph{negative-evidence constraints} encode unsuccessful edits as explicit “do-not-repeat” rules tied to file paths and change types, which prunes redundant actions in subsequent iterations.

\begin{figure}
  \centering
  \includegraphics[width=\linewidth]{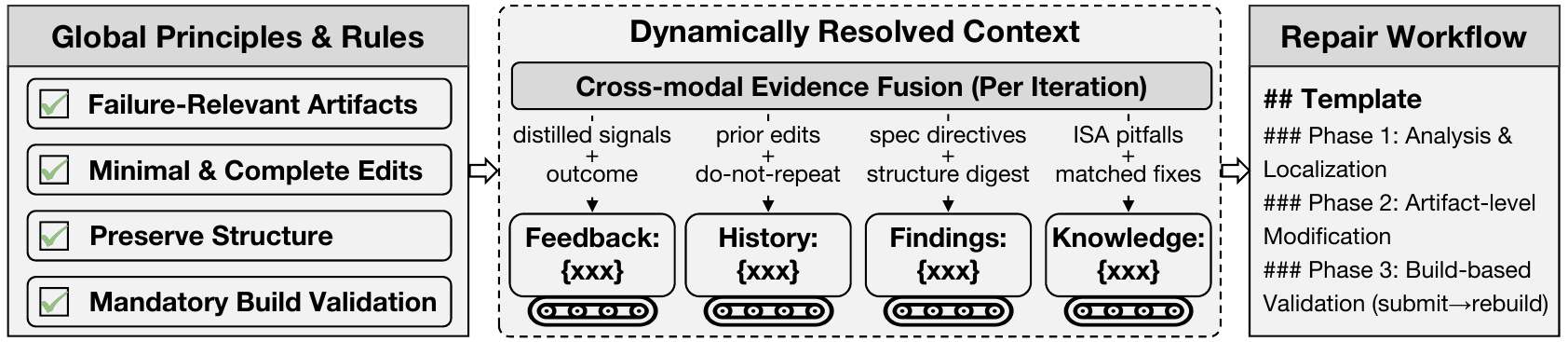}
  \caption{
  Compact prompt schema of \name{}. Per iteration, cross-modal evidence fusion organizes four evidence components into fixed prompt slots, combined with global repair rules and a build-validated workflow.}
  \label{fig:prompt}
\end{figure}

\subsubsection{Dynamic Prompt}
\label{sec:dynamic_prompt}

\name{} constructs a dynamic prompt before each repair iteration to translate iteration-aware evidence into more focused repair actions.
The prompt serves as the main control interface for the LLM agent that constrains how the LLM analyzes failures, applies artifact-level edits, and performs build-based validation.
Figure~\ref{fig:prompt} summarizes this schema, combining global repair rules with fixed evidence slots produced by cross-modal fusion and an explicit validation-driven workflow.

\noindent \textbf{Global Principles and Procedural Rules.}
The prompt encodes non-negotiable constraints that govern repair behavior.
These include evidence-guided prioritization of failure-relevant artifacts, enforcing minimal and complete edits, preserving original file structure and formatting, and requiring that all modifications be validated through the external build service.
In particular, each repair attempt must terminate with an explicit build submission, and unsuccessful validation automatically triggers the next iteration.
These rules prevent premature termination and ensure that all reported repairs are grounded in real build outcomes.

\noindent \textbf{Dynamically Resolved Context.}
At each iteration, the prompt instantiates four fused evidence slots (Feedback, History, Findings, and Knowledge), making the latest failure evidence, prior edits, and relevant contextual guidance explicit.
This slot-based organization reduces ambiguity from unstructured logs and supports history-aware reasoning across iterations.

\noindent \textbf{Structured Repair Workflow.}
To further reduce ambiguity in long-horizon reasoning, the prompt embeds a task-oriented workflow that decomposes repair into three ordered phases: failure analysis and localization, targeted artifact-level modification, and build-based validation.
Each phase is illustrated with lightweight examples and tool usage patterns, making the intended repair logic explicit.
This design mitigates unstructured exploration and encourages systematic, reproducible repair behavior across iterations.

\subsection{Analysis and Repair Orchestration}
\label{sec:orchestration}
Under the guidance of the Evidence-Preserving Repair Controller, this module executes the
analysis--repair stages of each iteration.
As illustrated in Figure~\ref{fig:framework}, it comprises two functional groups:
\emph{Failure Localization Tools} for the \emph{Analysis} phase and \emph{System-level Repair Actions}
for the \emph{Repair} phase.

\subsubsection{Failure Localization Tools}
To identify which artifacts are most likely responsible for the failure, \name{} provides three localization tools that summarize (1) the key error from the build log, (2) relevant dependency/configuration constraints from the recipe, and (3) a lightweight view of the package structure.

\noindent\textbf{Failure Signal Extraction.}

To transform voluminous and heterogeneous build logs into actionable repair guidance, \name{} employs a unified extraction tool that produces compact failure signals grounded in the latest build feedback. 
The tool follows a two-stage process.

First, it performs \emph{anomaly-focused log condensation} to isolate failure-relevant segments from background noise. 
This stage purifies the raw logs by removing redundant entries and replacing complex paths with concise placeholders. 
By cutting through this log noise, \name{} prevents the model from being overwhelmed and ensure it prioritizes salient failure evidence.
The extractor further applies phase-aware parsing to ensure that the distilled evidence remains strictly aligned with the temporal order of build stages such as \texttt{configure} or \texttt{install}. 
Following this pre-processing, the tool extracts contextual windows around diagnostic keywords (\eg \textit{undefined reference}, \textit{fatal error}) and normalizes them into abstract event templates using Drain~\cite{10.1145/3589334.3645442}. 
To ensure high recall, an LLM-based verifier classifies these candidate blocks to retain those containing explicit error patterns, which, together with their build-stage metadata, constitute the distilled failure signals.

Second, it performs \emph{retrieval-based contextualization} based on these distilled failure signals to query the controller-maintained \emph{Knowledge Context}.
The tool leverages the extracted templates, error keywords, and architecture hints within the signals to retrieve semantically related historical failures and ISA-aware knowledge via TF--IDF indexing and cosine similarity. 
This retrieved context complements the raw log evidence with recurring failure patterns and ecosystem conventions, thereby providing the domain expertise required for complex system-level repair.

\noindent \textbf{Dependency Constraints.}
Motivated by the findings in Section~\ref{sections/empirical_study}, \name{} employs a \emph{Dependency Constraints} tool to expose dependency and configuration-related constraints that frequently govern system-level build outcomes.
In Linux distributions, these constraints are primarily encoded in the package \emph{build recipe}, which specifies dependencies, macros, flags, and staged build commands; for RPM-based packaging, this recipe is implemented as a \texttt{.spec} file for RPM-based packaging to define dependencies, macros, and staged build commands.
The tool systematically parses metadata fields and macro definitions while segmenting build directives into discrete recipe stages such as \texttt{\%prep}, \texttt{\%build}, \texttt{\%install}, and \texttt{\%check}. 
By returning these extracted constraints in a structured JSON format, the tool enables the LLM to reason explicitly about ISA-dependent build requirements and configuration flaws. 
This approach provides structured evidence that eliminates the need for the agent to infer complex build logic implicitly from raw text.


\noindent \textbf{Structure Localization.}
To complement distilled failure signals and dependency constraints, \name{} employs a \emph{Structure Localization} tool that provides a global view of the package organization.
Given the package workspace, the tool traverses the directory tree, prunes non-essential artifacts such as documentation and generated files, and returns a hierarchical inventory of source files and configuration artifacts.
This structural evidence highlights plausible modification targets and cross-file relationships, thereby improving localization precision in complex, multi-language system-level packages.

\subsubsection{System-level Repair Actions}
After analysis localizes failure-relevant artifacts, \name{} enters the \emph{Repair} phase and applies system-level repair actions.
Unlike traditional program repair that focuses on source code alone, system-level package repair often requires coordinated updates across heterogeneous artifacts, including packaging recipes, configuration scripts, and compressed source archives.
Guided by the iteration-aware evidence context maintained by the controller, the orchestrator selects and applies one of the following repair actions, with all edits recorded for subsequent validation and history-aware reasoning.

\noindent\textbf{Archive Repackaging.}
System-level packages are commonly distributed as compressed source archives.
To support end-to-end repair, \name{} performs round-trip archive editing by unpacking the source archive into a temporary workspace, applying targeted edits, and reassembling the archive in the expected format for the build service.
To prevent invalid submissions, \name{} enforces two correctness constraints: the extracted workspace is never uploaded directly, and any modified content must be re-archived before validation.
This design ensures that each iteration is evaluated under realistic packaging conditions and remains compatible with production build workflows.

\noindent\textbf{Configuration Adaptation.}
\name{} supports configuration adaptation on the build recipe and related packaging scripts, including dependency declarations, macro definitions, compiler or linker flags, and stage-specific
directives.
Edits are applied through a full-file replacement policy to avoid partial or inconsistent changes, and each modification is recorded with its iteration index to support history-aware refinement.

\noindent\textbf{Source Modification.}
When configuration adaptation is insufficient, \name{} applies source modification to address code-level incompatibilities, missing definitions, or architecture- and toolchain-dependent assumptions.
The system first retrieves the complete content of a target file for inspection, then overwrites it with a revised version as a single atomic update.
For traceability, \name{} records a diff-style edit summary for each modification and appends it to the repair history, preventing redundant or conflicting edits in later iterations.

\subsubsection{Build-based Validation Tools}
As shown in Figure~\ref{fig:framework}, the \emph{Build-based Validation Tools} implement the \emph{Validation} phase of \name{} and close the repair loop by rebuilding the patched package under a reproducible environment and returning the resulting failure evidence to the controller.

\noindent\textbf{External Build Invocation.}
After a candidate fix is applied, \name{} submits the reconstructed package to the external build service through an upload-and-trigger interface.
To preserve a strict iteration boundary, each repair iteration issues exactly one build invocation, ensuring that all candidate fixes are evaluated under the same clean and reproducible conditions.

\noindent\textbf{Result-based Validation.}
During execution, \name{} monitors the submitted build until it reaches a terminal state.
To avoid indefinite waiting, the validator enforces a maximum monitoring window of 600 seconds, calibrated to typical build completion times observed in our package set.
If the build does not terminate within this window, \name{} marks the attempt as a timeout and treats it as an unsuccessful iteration.
A timeout differs from a normal failure in that the build service may not expose a complete new build log when the monitoring window expires.
In this case, the validator produces \emph{partial} feedback that records the timeout outcome and the latest observable build status returned by the build service.
For non-timeout failures, the validator retrieves the newly generated build log and stores it locally.
In both cases, the resulting build feedback is forwarded to the Evidence-Preserving Repair Controller to initiate the next iteration.

\noindent\textbf{Iterative Feedback Loop.}
By coupling external rebuilds with outcome-aware validation, \name{} forms a build-and-feedback loop.
Each iteration either terminates with a verified successful build or yields updated failure evidence that triggers the next round of localization and repair.

\section{Evaluation}
\label{sections/evaluation}


We evaluate \name{} on real-world system-level package build failures to answer the following research questions:
\begin{itemize}[leftmargin=*]
    \item \textbf{RQ3 (Effectiveness).} How effective is \name{} at repairing system-level package build failures?
    \item \textbf{RQ4 (Generalization).} Does \name{} generalize across different ISAs?
    \item \textbf{RQ5 (Iteration Necessity).} To what extent are iterative repair loops necessary?
    \item \textbf{RQ6 (Ablation).} How do orchestration-layer components affect repair performance?
    \item \textbf{RQ7 (Failure Analysis).} Why does \name{} fail on certain packages, and what limitations do these failures reveal?
\end{itemize}

\subsection{Experimental Setup}

\noindent\textbf{Dataset.}
We collect 219 RISC-V, 79 aarch64, and 83 x86\_64 build-failed packages from the Open Build Service (OBS).
These packages feature heterogeneous artifacts such as recipes, scripts, and source archives alongside multi-language code. 
Each package includes build logs capturing multi-stage failure evidence. 
We reproduce all reported failures before evaluation and use the RISC-V set as our primary benchmark, reserving others for ISA generalization.

\noindent\textbf{LLMs.}
We evaluate three LLMs to cover diverse deployment settings. 
\emph{GPT-5-mini} is selected for its balance of cost-efficiency and reliability in iterative tool-use scenarios. 
\emph{Qwen3-max} is chosen for its high-capacity orchestration of complex multi-step tasks , while \emph{DeepSeek-v3} serves as a powerful open-source representative with specialized code-oriented performance. 
All models operate in standard chat-generation mode without enabling separate reasoning variants, with the temperature set to 1.0 for all runs.


\noindent\textbf{Baselines.}
We compare \name{} against three classes of baselines to evaluate its relative effectiveness.
(1) \emph{Bare LLM.}
This baseline represents a non-agentic approach where the model proposes a repair in a single iteration from raw logs and the workspace. 
All auxiliary tools are disabled.
(2) \emph{Prior tool-augmented agents.}
We adapt Agentless~\cite{agentless_2025_fse}, CXXCrafter~\cite{CXXCrafter_2025}, and RepairAgent~\cite{RepairAgent_2025} to the system-level repair setting. 
Since these agents assume source-centric environments, we uniformly extend their inputs to package-level artifacts and removing their inherent restrictions to single-language codebases. 
We further replace their validation harnesses with our OBS build service.
Crucially, we preserve their original agentic workflows and internal tool-use logic to ensure a fair, reproducible comparison.
(3) \emph{\name{}(w/o evidence).}
To isolate the impact of our core principle, this variant employs the same tool suite as \name{} but clears the evidence after each iteration. 
It specifically evaluates whether tool access alone is sufficient for success without the coordination provided by an iteration-aware controller.



\noindent\textbf{Metric and Implementation.}
A repair is considered successful if the rebuilt package reaches the \emph{succeeded} state on OBS; we report the repair success rate as the fraction of successfully rebuilt packages.
We implement \name{} in Python and use GPT-5-mini as the default LLM.
Unless otherwise stated, we allow up to three repair iterations per package.
Experiments run on Ubuntu 22.04 with an Intel Xeon Gold 5416S CPU (64 cores) and 376~GB RAM.

\subsection{RQ3: Effectiveness of \name{}}
We evaluate the effectiveness of \name{} on 219 real-world RISC-V package build failures.
Table~\ref{tab:effectiveness} summarizes the repair outcomes, repair time, and build time of \name{} and all baselines.

\begin{table}
  \centering
  \captionsetup{font=small}
  \caption{Comparative evaluation of repair effectiveness and efficiency. 
  Success, Failed, and B/U (Broken/Unsolvable) counts partition the 219-package dataset. 
  B/U denotes cases with missing descriptions or unresolved dependencies. 
  \textit{Time-R} reports the average repair overhead per package (excluding the build phase),
while \textit{Time-B} reports the average external build duration in seconds. }
  \label{tab:effectiveness}
  \resizebox{\linewidth}{!}{
\begin{tabular}{l c c c c c c}
  \toprule
  \textbf{Methodology} & \textbf{Success} & \textbf{Failed} & \textbf{B/U} & \textbf{Success Rate (\%)} & \textbf{Time-R (s)} & \textbf{Time-B (s)} \\
  \midrule
  Bare LLM (Qwen3-max) & 2 & 190 & 27 & 0.91 & 48.53 & 149.78 \\
  Bare LLM (Deepseek-v3) & 3 & 177 & 39 & 1.37 & 44.56 & 134.00 \\
  Bare LLM (GPT-5-mini) & 4 & 191 & 24 & 1.83 & 40.38 & 106.75 \\
  \midrule
  Agentless~\cite{agentless_2025_fse} (GPT-5-mini) & 9 & 14 & 196 & 4.11 & 189.37 & 208.30 \\
  CXXCrafter~\cite{CXXCrafter_2025} (GPT-5-mini) & 45 & 41 & 133 & \textbf{20.55} & 303.05 & 526.22 \\
  RepairAgent~\cite{RepairAgent_2025} (GPT-5-mini) & 13 & 195 & 11 & 5.94 & 176.62 & 219.78 \\
  \name{}(w/o evidence) (GPT-5-mini) & 42 & 172 & 5 & 19.18 & 499.20 & 488.83 \\
  \midrule
  \name{} (Deepseek-v3) & 39 & 157 & 23 & 17.81 & 565.20 & 516.53 \\
  \name{} (Qwen3-max) & 57 & 99 & 63 & 26.03 & 507.82 & 549.78 \\
  \rowcolor{black!6} \name{} (GPT-5-mini) & \textbf{118} & 59 & 42 & \textbf{53.88} & 510.16 & 518.75 \\
  \bottomrule
\end{tabular}
  }
\end{table}

\noindent\textbf{Overall Repair Effectiveness.}
Using GPT-5-mini, \name{} repairs 118 out of 219 packages, achieving a success rate of \textbf{53.88\%}.
In contrast, bare LLM baselines remain below 2.00\% (0.91--1.83\%), indicating that single-pass log-to-patch generation without structured tool support rarely succeeds under build-service verification.

Among adapted prior agents, effectiveness remains limited.
Agentless repairs only 9 packages (4.11\%) and generates 196 \textit{B/U} cases. 
Its file-centric workflow struggles to (i) localize failure-relevant artifacts across complex packaging layouts, (ii) carry forward build feedback and repair history, and (iii) constrain edits under a fixed prompt schema with global rules. These limitations hinder actionable repairs, even after adapting inputs and OBS verification.
CXXCrafter achieves 20.55\% success. 
It focuses on generating and refining end-to-end build solutions such as Dockerfiles and scripts, rather than localizing and repairing package-level artifacts and archived sources.
Without history-aware constraints and evidence slots, repairs are limited to compilation symptoms and fail to address packaging and dependency-related issues.
RepairAgent repairs 13 packages (5.94\%) but lacks effective evidence organization. 
Failure evidence is not compacted into actionable signatures, unsuccessful edits are not flagged as “do-not-repeat,” and cross-artifact localization remains weak. 
These issues lead to drift and non-convergence under build-based feedback.
\name{}(w/o evidence) repairs 42 packages (19.18\%), showing improvement over bare LLMs. 
However, by discarding cross-iteration evidence, it repeats or introduces inconsistent actions, limiting convergence and further repair success.

\noindent\textbf{Time Efficiency.}
Table~\ref{tab:effectiveness} reports both repair-side overhead and OBS build duration.
Build time is largely determined by external verification and failure stage, while the main
runtime differences arise from the repair process itself.

Bare LLM baselines have the shortest repair time but almost never succeed, confirming that lightweight log-only reasoning is insufficient for system-level repair.
Among adapted agents, Agentless and RepairAgent exhibit relatively low repair time (189.37\,s and 176.62\,s).
This is consistent with their original design philosophy of keeping the agent workflow lightweight and file-centric, where the agent quickly selects a small set of candidate locations and generates a patch without maintaining iteration-level evidence slots or negative history.
CXXCrafter incurs higher repair time (303.05\,s), reflecting additional effort spent on
refining build procedures and environments.
\name{}(w/o evidence) and \name{} incur comparable repair time, suggesting that tool-driven analysis and artifact-level edits dominate the repair cost in this setting, while the additional overhead of iteration-aware evidence persistence is limited relative to the end-to-end build-and-verify loop.

\noindent\textbf{Impact of Failure Category and Software Languages.}
We further analyze repair effectiveness across the four failure categories identified in Section~\ref{sections/empirical_study}.
\name{} attains comparable success rates across categories: 68.18\% for test failures, 61.90\% for dependency failures, 55.56\% for compilation failures, and 50.00\% for packaging failures.
The reason is that category breakdown aligns with the types of evidence and repair targets exposed in package builds.
Dependency failures frequently require reasoning over build recipes and dependency constraints, while test failures are largely driven by test-stage evidence in build logs.
For compilation and packaging failures, directory structure and retrieved knowledge help map log evidence to concrete modification targets spanning recipes and source archives.
We also examine robustness across programming languages.
Among the 219 RISC-V packages, 80 involve multi-language mixtures, of which \name{} repairs 42.
For language-dominant packages, \name{} repairs 54/89 Python packages and 8/20 C/C++ packages, indicating that its repair decisions are primarily conditioned on package-level artifacts and build evidence rather than a single programming language.

\noindent\textbf{Impact of LLM Choice.}
Replacing GPT-5-mini with DeepSeek-v3 or Qwen3-max reduces \name{}’s success rate to 17.81\% and 26.03\%, respectively.
Despite these differences, \name{} consistently outperforms all baselines under each model, indicating that performance gains mainly come from iteration-aware evidence management and tool-orchestrated repair and validation, rather than relying on model capacity alone.

\begin{findingbox}
\textbf{Finding 3:} \name{} achieves a repair success rate of \textbf{53.88\%} using GPT-5-mini, outperforming the strongest baselines.
This improvement holds consistently across failure categories (50.00–68.18\%) and language settings, including 52.50\% success on multi-language packages.
\end{findingbox}

\begin{table}
  \centering
  \captionsetup{font=small}
  \caption{Preliminary results of \name{}'s generalizability across different Instruction Set Architectures (ISAs).}
  \label{tab:generalization}

  \resizebox{\linewidth}{!}{
  \begin{tabular}{l l c c c c c c}
    \toprule
    \textbf{ISA} & \textbf{Methodology} &
    \textbf{Success} & \textbf{Failed} & \textbf{B/U} &
    \textbf{Success Rate (\%)} &
    \textbf{Time-R (s)} & \textbf{Time-B (s)} \\    
    \midrule

    \multirow{5}{*}{aarch64}
      & Agentless~\cite{agentless_2025_fse} (GPT-5-mini)        & 1  & 67 & 11 & 1.27  & 197.42 & 142.01 \\
      & CXXCrafter~\cite{CXXCrafter_2025} (GPT-5-mini)       & 15 & 9  & 55 & 18.99 & 216.99 & 318.31 \\
      & RepairAgent~\cite{RepairAgent_2025} (GPT-5-mini)      & 4  & 64 & 11 & 5.06  & 212.07 & 176.44 \\
      & \name{}(w/o evidence) (GPT-5-mini) & 22 & 43 & 14 & 27.85  & 550.20  &  333.76 \\
      \rowcolor{black!6}  & \name{} (GPT-5-mini)  & \textbf{33} & 40 & 6  & \textbf{41.77} & 547.63 & 466.21 \\
    \midrule

    \multirow{5}{*}{x86\_64}
      & Agentless~\cite{agentless_2025_fse} (GPT-5-mini)        & 3  & 53 & 27 & 3.61  & 181.09 & 219.25 \\
      & CXXCrafter~\cite{CXXCrafter_2025} (GPT-5-mini)       & 9  & 12 & 62 & 10.84 & 208.24 & 317.40 \\
      & RepairAgent~\cite{RepairAgent_2025} (GPT-5-mini)      & 1  & 70 & 12 & 1.20  & 210.72 & 139.29 \\
      & \name{}(w/o evidence) (GPT-5-mini) & 19 & 51 & 13 & 22.89  &  663.47  & 347.00 \\
      \rowcolor{black!6}  & \name{} (GPT-5-mini)          & \textbf{39} & 36 & 8  & \textbf{46.99} & 525.41 & 504.37 \\
    \bottomrule
  \end{tabular}}
\end{table}

\subsection{RQ4: Generalization across Different ISAs}
To evaluate the architectural robustness of \name{}, we investigate its ability to generalize beyond the RISC-V ecosystem. 
A pivotal design principle of \name{} is the separation of concerns, which decouples the ISA-agnostic repair workflow from the ISA-specific \emph{Knowledge Context}. 
This architecture ensures that all core modules operate on generic system-level artifacts, including build logs, packaging recipes, and source archives, without being hard-coded for any specific ISA.

\noindent\textbf{Seamless Adaptation.} 
We instantiated \name{} for x86\_64 and aarch64 by simply swapping the underlying architecture-related data sources in the \emph{Knowledge Context}, as detailed in Section~\ref{sec:context-manager}. 
As summarized in Table~\ref{tab:generalization}, \name{} achieves a repair success rate of 41.77\% (33/79) on aarch64 and 46.99\% (39/83) on x86\_64. 
The consistency of these results across three distinct ISAs demonstrates that our evidence-preserving principle serves as a universal solution for system-level failures.

\noindent\textbf{Evidence Preservation and Baseline Comparison.} 
\name{} consistently outperforms all adapted agent baselines across both architectures by a significant margin. 
Specifically, Agentless achieves only 1.27\% on aarch64 and 3.61\% on x86\_64, whereas CXXCrafter and RepairAgent fail to exceed 19\% and 6\% respectively. 
This performance gap indicates that tool access alone is insufficient for navigating heterogeneous package ecosystems. 
Overall, these results show that without structured evidence management, generic agents struggle to interpret and reuse complex failure signals.

\begin{findingbox}
\textbf{Finding 4:} \name{} repairs 41.77\% of aarch64 and 46.99\% of x86\_64 packages by updating only the ISA-specific Knowledge Context.

\end{findingbox}

\begin{figure}
  \centering
  \includegraphics[width=0.95\linewidth]{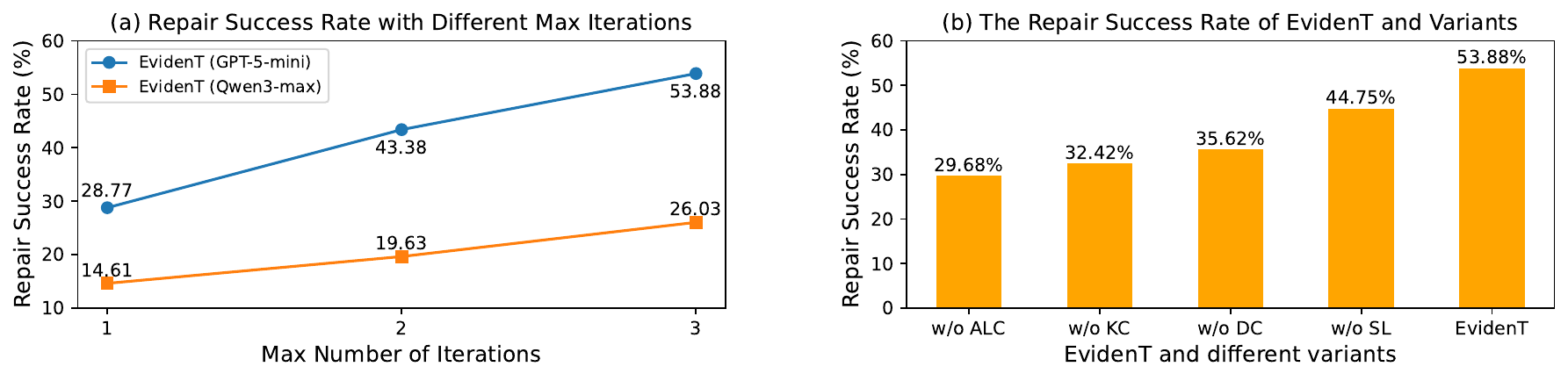}
  \caption{Repair success rates of \name{}. 
    (a) Repair success rates under different maximum iteration budgets (1--3) for GPT-5-mini and Qwen3-max. 
    (b) Success rates of ablated variants removing each component.
    }
  \label{fig:repair success rate with and variants}
\end{figure}

\subsection{RQ5: Iteration Necessity}
System-level package repair relies on build-based verification, where each repair attempt is assessed by an external rebuild and the next decision depends on the latest build feedback and prior edits.
For each iteration, the Evidence-Preserving Repair Controller refreshes \emph{Build Feedback}, \emph{Repair History}, and \emph{Cached Findings} (Section~\ref{sec:context-manager}) before the subsequent analysis and repair steps.
Figure~\ref{fig:repair success rate with and variants}(a) reports repair success rates as the maximum number of iterations increases from one to three.
With one iteration, \name{} repairs 63 packages (28.77\%) using GPT-5-mini and 32 packages (14.61\%) using Qwen3-max.
With two iterations, the numbers increase to 95 (43.38\%) and 43 (19.63\%), respectively.
With three iterations, repair success further increases to 118 (53.88\%) and 57 (26.03\%).
These results indicate that additional iterations improve repair effectiveness under build-service feedback.

\noindent\textbf{Repair Outcome Patterns.}
We categorize repair trajectories by the iteration at which they converge.
First-iteration successes typically occur when the initial build feedback contains a localized and actionable error signal that can be mapped to a single repair target, such as a build recipe directive or a specific source file.
Second-iteration successes often arise when the first attempt addresses part of the failure, and the subsequent rebuild exposes refined evidence, such as an error that shifts to a later build stage or a newly surfaced missing dependency, enabling a more targeted follow-up edit.
Non-convergent cases persist when iterative rebuilds fail to provide refinement signals.
In particular, some packages terminate early due to corrupted archives, missing build recipes, or invalid build metadata, leaving no reliable target for modification.
Other cases produce inconsistent failure signatures across iterations, so updates cannot be narrowed to a stable artifact or stage.

\begin{findingbox}
    \textbf{Finding 5:} Increasing the iteration budget from one to three raises repair success by 25.11 percentage points with GPT-5-mini (28.77\%\,$\rightarrow$\,53.88\%) and by 11.42 points with Qwen3-max (14.61\%\,$\rightarrow$\,26.03\%).
\end{findingbox}

\subsection{RQ6: Contribution of Analysis \& Repair Orchestration Components}
To evaluate the necessity of the evidence-preservation principle, we perform an ablation study by removing one key component at a time.
We compare the full \name{} against four variants: (i) $\text{\name{}}_{\text{w/o ALC}}$, which disables Anomaly-focused Log Condensation; (ii) $\text{\name{}}_{\text{w/o KC}}$, removing Knowledge Context retrieval; (iii) $\text{\name{}}_{\text{w/o DC}}$, omitting Dependency Constraints; and (iv) $\text{\name{}}_{\text{w/o SL}}$, bypassing Structure Localization.

\noindent\textbf{$\text{\name{}}_{\text{w/o ALC}}$.}
Disabling ALC causes the most significant performance regression, with the success rate plummeting from 53.88\% to 29.68\%. 
Without this stage, build feedback is no longer distilled into compact, phase-aligned failure signatures.
Therefore, subsequent analysis cannot reliably prioritize salient signals from noisy logs and becomes more prone to leading to misaligned and ineffective edits.

\noindent\textbf{$\text{\name{}}_{\text{w/o KC}}$.}
Removing \emph{Knowledge Context} retrieval reduces the success rate to 32.42\%.
This mainly affects cases where the repair depends on ISA or toolchain-specific conventions not explicit in the current build log.
The agent can no longer leverage matched historical experience, facing higher ambiguity.

\noindent\textbf{$\text{\name{}}_{\text{w/o DC}}$.}
The absence of Dependency Constraints lowers the success rate to 35.62\%.
Without structured recipe-level constraints, the agent must recover dependencies, macros, and stage directives from raw text, which increases incomplete or inconsistent configuration edits.

\noindent\textbf{$\text{\name{}}_{\text{w/o SL}}$.}
Removing Structure Localization yields a smaller but consistent decrease to 44.75\%.
The structural inventory helps map localized signals to concrete edit candidates in large workspaces.
Without it, target selection becomes less stable for multi-component packages.


\begin{findingbox}
    \textbf{Finding 6:} Removing any component reduces repair success from 53.88\% to 29.68--44.75\%, indicating that effective repair in \name{} relies on their complementary contributions.
\end{findingbox}


\subsection{RQ7: Failure Analysis and Limitations}
\noindent\textbf{Tool Usage Behavior.}
We analyze 7,923 tool invocations across 219 packages.
Tool calls largely follow the intended analysis--repair--validation workflow, with only three cases exhibiting misordered execution.
For 34 packages, the repair process terminates early before completing required validation steps, typically due to corrupted archives, missing build recipes, or invalid build metadata.
Overall, unsuccessful repairs are more often associated with insufficient or unusable failure evidence than with systematic tool misuse.

\noindent\textbf{Limitations.}
\name{} is most effective when build feedback provides stable and actionable refinement signals across iterations.
Some packages remain unrepaired when the build service cannot produce complete evidence (e.g., incomplete artifacts or unstable failure traces), preventing the controller from conditioning subsequent iterations on comparable feedback.
Handling such cases may require additional recovery support for invalid inputs and partial build feedback.

\begin{findingbox}
\textbf{Finding 7:} Among 7,923 tool invocations, misordered execution is rare (3 cases), while 34 packages fail due to early termination under invalid or incomplete build artifacts.
\end{findingbox}

\subsection{Case Study}
We present a case study on \texttt{postquantumcryptoengine}~\cite{openSUSE_postquantumcryptoengine}, a C++ library for post-quantum cryptography, to illustrate how \name{} performs evidence-driven
system-level repair under complex artifact and dependency conditions.
The initial build on \texttt{riscv64} fails during compilation with unresolved template symbols and missing type definitions related to the Open Quantum Safe (OQS) library.

Using \emph{Failure Signal Extraction}, \name{} condenses the raw build log into a small set of architecture-tagged error signatures, which indicate that the failure originates from C++ header processing rather than linker configuration.
In parallel, \emph{Dependency Constraints} analysis over the spec file exposes strict version and feature requirements on \texttt{liboqs}, ruling out missing-package installation as the primary cause.
To identify the concrete faulty artifact, \name{} applies \emph{Structure Localization} to inspect the package workspace and source archive layout. 
This step narrows the failure to a truncated header file, \texttt{crypto.hh}, whose class declarations are incomplete and lack required OQS includes, preventing correct instantiation of cryptographic components.
Guided by this localized evidence, \name{} reconstructs the missing class hierarchy and restores the expected OQS interfaces through a targeted source-level modification.
After repackaging the modified archive, \name{} validates the fix through the external build service.
Two intermediate iterations expose residual compilation errors, which are resolved by refining the header definitions based on updated build feedback.
The third iteration completes successfully, yielding a clean rebuild.

\section{Discussion}

\noindent\textbf{Design Trade-offs and Remaining Failure Modes.}
\name{} prioritizes evidence-driven and minimally repairs to balance immediate success with long-term maintainability, as aggressive modifications often undermine build reproducibility. 
For some unresolved cases, failures are often constrained by ecosystem-level factors rather than improper tool orchestration alone.
In particular, incomplete or corrupted source artifacts, unstable dependency availability, and non-deterministic build or test behaviors may prevent convergence even when failure localization is accurate.
These observations suggest that repair efficacy is bounded by both agent reasoning and the quality of the surrounding build infrastructure.

\noindent\textbf{Build Success vs. Functional Correctness.}
A successful build does not necessarily guarantee full functional correctness at runtime.
Our evaluation focuses on build-level success as a practical and reproducible proxy for large-scale system-level repair.
While this metric captures the ability to resolve concrete build failures, additional validation (\eg regression testing, performance benchmarking, and integration testing) is required to assess post-build functionality.
Incorporating systematic runtime testing into \name{} remains an important direction for future work.

\noindent\textbf{Threats to Validity.}
Several factors may affect the validity of our results.
First, repair performance can vary across LLMs due to differences in tool-use reliability and code generation capability, as well as the inherent stochasticity of model outputs.
To mitigate this effect, we repeat all experiments twice using the default model (GPT-5-mini) and observe no noticeable variation in repair success.
Second, our evaluation relies on the Open Build Service (OBS) for build-based validation, whose updates or instability may affect consistency.
We plan to address this limitation by adopting containerized build environments (\eg Docker) to further improve reproducibility.


\section{Conclusion}
In this paper, we present \name{}, an evidence-preserving framework for iterative system-level package repair.
We conduct an empirical study of real-world package build failures to characterize their diverse root causes across heterogeneous artifacts, build stages, and validation environments.
Based on these observations, \name{} integrates coordinated tool support for failure analysis, artifact-level repair, and build-based validation, enabling repairs to be guided by accumulated evidence over multiple iterations.
Our evaluation on real package failures across multiple instruction set architectures demonstrates the effectiveness and portability of \name{}.
Specifically, \name{} repairs 118/219 packages on RISC-V (53.88\%), 33/79 on aarch64 (41.77\%), and 39/83 on x86\_64 (46.99\%), substantially outperforming bare LLM baselines and representative agent-based approaches.
These results highlight the importance of preserving and reusing failure evidence, as well as validating repairs through external build services, for scalable automated package maintenance.

\section{Data Availability}

We have made our source code and datasets publicly available through an anonymous repository at \url{https://anonymous.4open.science/r/EvidenT-1638/README.md}.

\bibliographystyle{ACM-Reference-Format}
\bibliography{main}






\end{document}